\documentclass{emulateapj}
\shorttitle{Dust in SN2006jc}
\shortauthors{Smith et al.}
\begin{document}


\title{Dust Formation and H\lowercase{e}~II~$\lambda$4686 emission in
the Dense Shell of the Peculiar Type~I\lowercase{b}
Supernova~2006\lowercase{jc}}

\author{Nathan Smith, Ryan J.\ Foley, \& Alexei V.\ Filippenko}

\affil{Department of Astronomy, University of California, 
Berkeley, CA 94720-3411}

\begin{abstract}

We present evidence for the formation of dust grains in an unusual
Type~Ib supernova (SN) based on late-time spectra of SN~2006jc.  The
progenitor suffered an outburst qualitatively similar to those of
luminous blue variables (LBVs) just 2 yr prior to the SN, and we
propose that the dust formation is a consequence of the SN blast wave
overtaking that LBV-like shell.  The key evidence for dust formation
is (a) the appearance of a red/near-infrared continuum emission source
that can be fit by $T \approx 1600$~K graphite grains, and (b) fading
of the redshifted sides of intermediate-width He~{\sc i} emission
lines, yielding progressively more asymmetric blueshifted lines as
dust obscures receding material.  This provides the strongest case yet
for dust formation in any SN~Ib/c.  Both developments occurred between
51 and 75~d after peak brightness, while the few other SNe observed to
form dust did so after a few hundred days.  Geometric considerations
indicate that dust formed in the dense swept-up shell between the
forward and reverse shocks, and not in the freely expanding SN ejecta.
The rapid cooling leading to dust formation may have been aided by
extremely high shell densities of 10$^{10}$ cm$^{-3}$, indicated by
He~{\sc i} line ratios.  The brief epoch of dust formation is
accompanied by He~{\sc ii} $\lambda$4686 emission and enhanced X-ray
emission, suggesting a common link.  These clues suggest that the
unusual dust formation in this object was not attributable to
properties of the SN itself, but instead --- like most peculiarities
of SN~2006jc --- was a consequence of interaction with the dense
environment created by an LBV-like eruption 2~yr before the SN.

\end{abstract}


\section{INTRODUCTION}

Supernova (SN) 2006jc is a peculiar Type Ib event that challenges our
understanding of the late-time evolution of massive stars in several
important ways.  It is one of only three known SNe~Ib/c (see
Filippenko 1997 for a review of supernova spectral types) that exhibit
strong and relatively narrow (widths of $\sim$10$^3$ km s$^{-1}$;
hereafter referred to as ``intermediate width'') emission
lines of He~{\sc i} in their optical spectra (Foley et al.\ 2007;
Pastorello et al.\ 2007) --- the other two known examples being
SN~1999cq (Matheson et al.\ 2000) and SN~2002ao (Foley et al.\ 2007).
SNe~Ib/c probably mark the deaths of Wolf-Rayet (WR) stars, but the
intermediate-width He~{\sc i} lines in these three objects indicate much denser
circumstellar material (CSM) than that around all other SNe~Ib/c, and
similarly, inferred progenitor mass-loss rates much higher than those
typically observed in WR stars (Crowther 2007). X-rays detected from
SN~2006jc also imply strong CSM interaction (Immler et al.\ 2008).

What may turn out to be the most important aspect of SN~2006jc,
unique among all SNe observed so far, is that the
progenitor star was observed to have suffered a giant outburst
qualitatively similar to those of luminous blue variables (LBVs) just
2~yr before being discovered as a SN (Nakano et al.\ 2006; Pastorello
et al.\ 2007).  These outbursts, sometimes called ``supernova
impostors'' when they are seen in other galaxies, are thought to be
analogous to the 19th-century eruption of $\eta$~Carinae, which
ejected $\sim$10~M$_{\odot}$ within about a decade (Smith et al.\ 2003).
The best-known extragalactic example is SN~1961V in NGC~1058 (e.g.,
Goodrich et al.\ 1989; Filippenko et al.\ 1995; Van Dyk et al.\ 2002).

LBVs are usually presumed to be stars in transition from core-H
burning to He burning, so they should not explode as SNe for another
few hundred thousand years. However, there is mounting evidence to the
contrary, that some LBVs --- or related H-rich stars with similar
eruptive mass loss --- defy expectations of stellar evolution models
and might in fact explode as SNe (Smith \& Owocki 2006; Kotak \& Vink
2006; Gal-Yam et al.\ 2007; Smith 2007; Smith et al.\ 2007).  The
apparent magnitude of the transient event 2 yr before SN~2006jc was
consistent with one of these giant LBV eruptions, but to see one from
a WR star is highly unexpected, as noted by Foley et al.\ (2007) and
Pastorello et al.\ (2007).

Stellar evolution theory currently makes no prediction that the
LBV-like instability should continue to play a role through the start
of the WR phase\footnote{To be fair, though, stellar evolution theory
makes no clear prediction of the LBV instability during the LBV phase
either, since the mechanism is not understood.}, but there are some
observational indications that the two phenomena may be closely
related (see Smith \& Conti 2008).  It is an observational fact that
giant LBV eruptions can remove large amounts of mass (of order 
10~M$_{\odot}$) in short bursts that last a decade, occurring
multiple times, and that these sudden mass ejections may therefore be
the dominant mechanism responsible for shedding a star's H-rich
envelope to produce a WR star (Smith \& Owocki 2006).  In retrospect,
perhaps it should not be surprising, then, if the instability
responsible for making a WR star persists into the early WR phase
itself.

A few well-known examples point in this direction.  Among the more
luminous and most exemplary of the LBVs are AG~Carinae in our Galaxy
and R~127 in the Large Magellanic Cloud.  Both of these LBVs are surrounded 
by massive ring nebulae (Stahl 1987), and both have been classified as
Ofpe/WN9 stars in their hot quiescent phases when they were not
exhibiting the LBV behavior (Stahl 1986; Stahl et al.\ 1983).  The
Ofpe/WN9 stars (Bohannan \& Walborn, 1989; Crowther et al.\ 1995;
alternatively called WN11) are probably not core-He burning WR stars,
and many have never been observed to undergo an LBV eruption; they are
still H-rich, but they are thought to represent one of the
transitional phases between O stars and WR stars.  

Another intriguing object is the eclipsing binary HD~5980
(Koenigsberger 2004), the most luminous star known in the Small Magellanic
Cloud.  It consists of two stars with WN-like spectra, one of which
suffered a brief LBV-like outburst in 1994 (Barb\'{a} et al.\ 1995;
Koenigsberger 2004).  The outburst was shorter in duration, less
luminous, and ejected far less mass than the eruptions of $\eta$~Car
or SN~1961V, but it had properties in common with some LBV outbursts.
The erupting star was H rich (Koenigsberger 2004; Koenigsberger et
al.\ 2000; Moffat et al.\ 1998), so although its spectrum showed
WN-like features, it too has not yet shed its H envelope to become a
{\it bona fide} WR star and is probably instead a transition object
(Smith \& Conti 2008).  Nevertheless, these connections may be
relevant to SN~2006jc after all, which had weak H$\alpha$ emission in
its spectrum (Foley et al.\ 2007), although still H poor.

In a previous paper (Foley et al.\ 2007, hereafter Paper~I) we
presented and discussed our early-time optical spectra and light
curves of SN~2006jc. The most important result for our purposes here
is that the unusual intermediate-width He~{\sc i} lines of SN~2006jc
are probably produced through ejecta interacting with its dense CSM
that was created by the LBV-like eruption observed 2~yr before the SN.
The implications for stellar evolution are quite provocative, as noted
above, because this is the first case of an LBV-like eruption in a
H-poor star.  Pastorello et al.\ (2007) reached a similar conclusion
independently, although they also propose that a classical LBV
outburst may have occurred in a companion star in a binary system
instead of the same star that died to make SN~2006jc.  However, the
observed emission from the CSM of SN~2006jc requires that the
companion's LBV outburst would need to have been H poor as well,
making this scenario doubtful.

Here we explore the influence of this dense CSM created in the
LBV-style outburst from the perspective of evidence for the rapid
formation of dust grains in SN~2006jc.  Based on near-infrared
(near-IR) broad-band photometry, Arkharov et al.\ (2006) found that
SN~2006jc was rebrightening in the $H$ and $K$ bands in late-November
through early-December 2006, and they conjectured that this unusual
outburst was likely to have ``some kind of relation with dust.''
Whether the dust was pre-existing or formed in the SN is not obvious
from the near-IR photometry alone.

Grain formation would be very unusual for SNe~Ib/c; the Type Ib
SN~1990I is the only previous case where dust formation is thought to
have occurred (Elmhamdi et al.\ 2004).  In fact, clear evidence for
dust formation in SNe has been elusive for any type of SN, even though
the idea that SNe could form dust was postulated long ago (e.g.,
Cernuschi, Marsicano, \& Codina 1967; Hoyle \& Wickramasinghe 1970).
Observations of SN~1979C (Merrill 1980) and SN~1980K (Dwek et al.\
1983) showed suggestive evidence for dust formation, but SN~1987A
provided the first well-established case.  SN~1987A showed an IR
excess, a simultaneous decrease in its optical light, and a systematic
blueshift in its broad emission lines formed in the SN ejecta
(Danziger et al.\ 1989; Lucy et al.\ 1989; Gehrz \& Ney 1989; Wooden
et al.\ 1993; Colgan et al.\ 1994; Wang et al.\ 1996; Moseley et al.\
1989; Dwek et al.\ 1992).  Two more-recent cases providing evidence
for dust formation are SN~1999em (Elmhamdi et al.\ 2003) and SN~2003gd
(Sugerman et al.\ 2006), although Meikle et al. (2007) found that the
mass of dust formed in SN~2003gd was very small.  All three of these
were Type II explosions where the dust formed in the SN ejecta.  Dust
in SN remnants (SNRs) is usually dominated by swept-up interstellar or
circumstellar dust.  A rare exception is the Crab Nebula, where
$\lesssim$0.01 M$_{\odot}$ of dust is seen (Temim et al.\ 2006).

An observed near-IR excess alone is not sufficient evidence for dust
formation.  Unlike the case of classical novae, where the temporal
development of the observed excess shows clear evidence for new grain
formation in the ejecta (e.g., Gehrz 1988), in SNe the observed IR
excess is often attributed to a light echo (Dwek 1983; Emmering \&
Chevalier 1988; Bode \& Evans 1980).  This arises when light from the
SN illuminates and heats pre-existing dust in the CSM.

In this paper we present evidence from optical spectra that the
near-IR rebrightening of SN~2006jc noted first by Arkharov et al.\
(2006) is indeed caused by dust {\it formation}, and not an IR light
echo.  Our key result is that the dust formation in SN~2006jc is
directly linked to its unusually dense environment, and that dust
formed in the dense post-shock shell rather than within freely
expanding SN ejecta. SN~2006jc is the first clear example of a SN
forming dust in this way.

In \S 2 we present new late-time optical spectra of SN~2006jc that were
obtained after those published in Paper~I, and in \S 3 and \S 4 we
discuss two independent lines of evidence that dust has formed.  In 
\S 5 we give reasons why the near-IR excess in SN~2006jc is probably
{\it not} the result of pre-existing circumstellar dust heated by the
SN.  In \S 6 we consider the geometry and location of this dust, and
in \S 7 we summarize our key results.

\section{NEW OPTICAL SPECTRA OF SN~2006\lowercase{jc}}

In Paper~I, we presented and discussed early-time low-resolution and
moderate-resolution spectra of SN~2006jc up to day 44 after the time
when photometric monitoring began (near the $B$-band peak).  These
were obtained with the Kast spectrograph (Miller \& Stone 1993) on the
Lick Observatory 3-m Shane reflector, as well as with the LRIS (Oke et
al.\ 1995) and DEIMOS (Faber et al.\ 2003) spectrographs mounted on
the 10-m Keck I and II telescopes, respectively.  Here we present
additional optical spectra obtained at later times using the same
observational setups and data-reduction techniques (see Paper I for
further details).  The
additional spectra discussed here were obtained on UT dates of 2006
December 1.4 (day 51; Lick Kast), 2006 December 25.5 (day 75; Keck LRIS), 
2007 January 21.5 (day 102; Keck LRIS), and 2007 February 16.5 (day 128;
Keck DEIMOS).

\section{EVIDENCE FOR DUST: THE RED CONTINUUM}

\subsection{Changes in the Continuum Shape}

Figure 1 shows optical spectra obtained during the first $\sim$100~d
of SN~2006jc, with the flux scaled to highlight changes in the
relative continuum shape (normalized roughly to the $R$-band flux).
As noted in Paper~I, there are pronounced changes with time in the
relative strengths of intermediate-width features like the He~{\sc i}
lines, as well as the broad features emitted by the SN ejecta such as
the Ca~{\sc ii} near-IR triplet and O~{\sc i} $\lambda$7774.

\begin{figure}
\epsscale{0.99}
\plotone{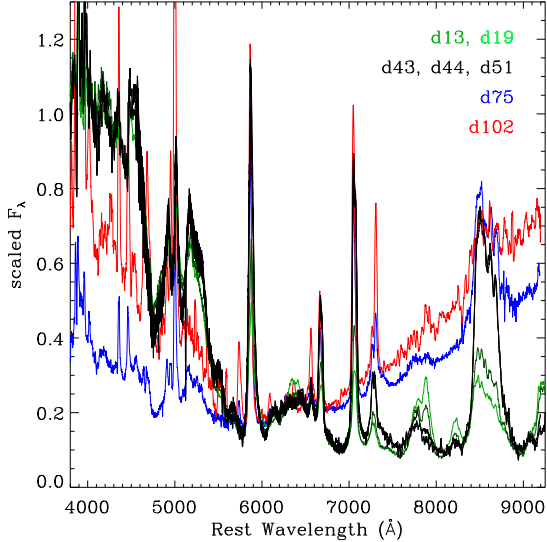}
\caption{Optical spectra of SN~2006jc from early times (two weeks
  after explosion) until late times, scaled to highlight changes in
  the relative continuum shape. By late-December 2006 (day 75) through
  late-January 2007 (day 102), SN~2006jc showed a pronounced red
  continuum that had not been present before.  Its strength was
  comparable to that of the blue emission, creating a ``U''-like shape
  in the optical continuum.  The spectra were normalized near 6400
  \AA.}
\end{figure}

Setting these contributions from specific lines aside, though, it is
stunning how little the underlying shape of the optical spectrum
changes at early times.  Until at least $\sim$50~d (green and black
in Fig.\ 1), there is essentially {\it no change} in the shape of the
continuum to within a few percent, despite the rapid fading of the SN
by a factor of $\sim$15 during the same time interval.  The photometric 
$B-R$ color stays nearly constant during this time as well (Paper I).  The
spectrum persistently shows a flat, weak, red continuum, but shortward
of $\sim$5500~\AA, the continuum abruptly jumps upward to produce a
strong blue/near-UV continuum.  Even the detailed undulations of this
blue ``continuum'' did not change by more than a few percent during
the first $\sim$50~d (Fig.\ 1).  This steep blue continuum remains
puzzling, but was also seen in the Type IIn SN 1998S (Leonard et al. 
2000); in Paper~I we suggested that it may be caused by many
blended Fe lines, perhaps due to fluorescence (broadly construed to
include UV or collisional excitation). We noted that the two other 
SNe~Ib/c with strong, intermediate-width He~{\sc i} lines (SNe 2002ao 
and 1999cq) show similar blue continua, although not as strong as in 
SN~2006jc, perhaps due to line-of-sight reddening from dust.

The continuum shape of SN~2006jc then took a bizarre turn sometime in
mid-December 2006.  By Dec. 25 (day 75; blue in Fig.\ 1) the continuum
looked completely different.  The blue-wavelength continuum was still
present, although it was not as steep, having dropped by almost a
factor of three relative to 6500~\AA.  The most striking change was at
red and near-IR wavelengths, where SN~2006jc developed a strong
continuum component not present in any earlier spectra.  With the
spectrum rising into the blue {\it and} rising sharply into the
near-IR, the optical continuum took on a ``U''-like shape that has not
been documented as clearly in any other SN.  This ``U''-shaped
continuum was still present a month later in late-January 2007 (day
102; red in Fig.\ 1).

Figure 2 shows our late-time optical spectra of SN~2006jc, with the
last four epochs plotted individually.  The first epoch at day 51 is
essentially identical to the day 43/44 spectra that were the latest
ones discussed in Paper~I.  The spectra at days 75 and 102 clearly
show the new near-IR excess emission that subsequently disappeared by
day 128.

\begin{figure}
\epsscale{0.99}
\plotone{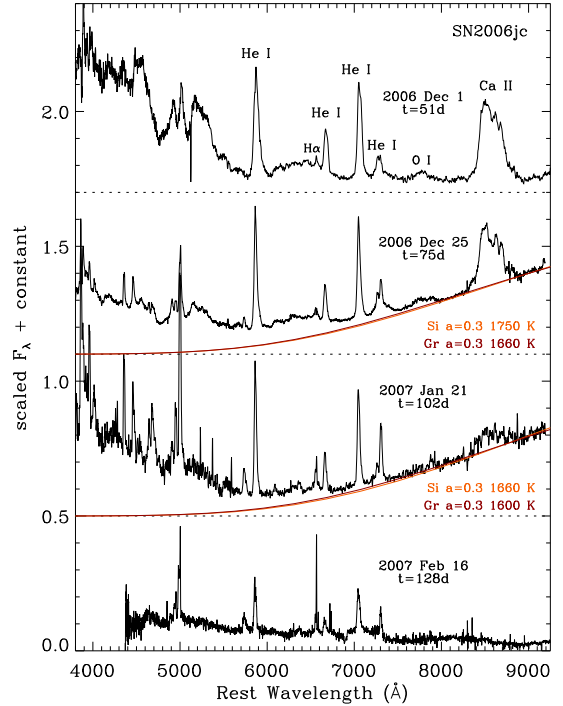}
\caption{Late-time optical spectra of SN~2006jc from Figure 1 but
  plotted individually.  The pronounced red continuum that was present
  in late-December 2006 through January 2007 had not been present
  before, and disappeared by mid-February 2007. (We suspect that our
  final spectrum on day 128 may be contaminated by an underlying star
  cluster and H~{\sc ii} region; prominent, very narrow H$\alpha$
  emission is present.)  At the epochs when the strong red continuum
  is present, red and orange curves show models for emission from
  graphite and silicate grains, respectively, at the temperatures
  indicated.  The curves shown here use emissivities for silicate and
  graphite grains of radius $a = 0.3$~$\micron$ from Draine \& Lee
  (1984), but the results do not depend strongly on assumed grain
  radius as long as $a$ is less than $\sim$1~$\micron$.}
\end{figure}

The most plausible explanation for this transient red/near-IR
continuum emission is that it is caused by hot dust -- either newly
formed hot dust, or pre-existing dust in the CSM that is heated by
energy input from the SN.  We will return to these options later.  For
now, the purpose of Figure 2 is simply to illustrate that hot dust can
account for the transient red continuum shape.

Although the shape of the red continuum excess can be approximately
fit by a $\sim$2000~K Planck function, optically thin emitting dust
grains will have wavelength-dependent emissivity.  As long as the
grains are not larger than about $a = 1$~$\micron$, the dust will
typically have emissivity proportional to $\lambda^{\beta}$, with
$-2\la\beta\la -1$.  The red and orange curves in Figure 2 are plotted
adopting the emissivities of graphite and astronomical silicate,
respectively, both with $a = 0.3$~$\micron$ (Draine \& Lee 1984).
They are nearly identical, meaning that from the emission at
these wavelengths alone, we are unable to directly constrain the grain
chemistry.  However, the high temperature needed to produce the red
continuum shape gives an important clue.  With these emissivities, the
grain temperatures are around 1600~K for graphite particles, or
slightly warmer for silicates (Fig.\ 2).  Standard theory for carbon
condensation (Clayton 1979) and models of dust formation in SN ejecta
(Todini \& Ferrara 2001) predict that nucleation starts at $T \approx
1800$~K for carbon grains, while silicates require lower temperatures
of 1000--1200~K, as typically seen in novae (Gehrz 1988).  Thus,
Figure 2 is compatible with the notion that this red emission arises
from freshly synthesized, amorphous carbon dust in SN~2006jc, while the
high temperatures seem problematic for silicates.

\subsection{Luminosity and Mass of the Hot Dust}

Figure 3 shows the same data and the same silicate and graphite dust
models for day 75 from Figure 2, but plotted as $\lambda$F$_{\lambda}$
on an absolute flux scale with a broader wavelength range.  We would
normally use our optical photometry of SN~2006jc to directly check our
absolute flux calibration of the spectrum, but our optical photometry
of SN~2006jc ended just over a week before the day 75 spectrum (black
in Fig.\ 3) was obtained.  Extrapolating our light curve using the
late-time decline rate of 0.066 mag d$^{-1}$ (Paper I), we estimate
SN~2006jc to have $R = 18.8$ mag on 2006 Dec. 25.  Using this as a
guide, we derive a continuum flux at 6900~\AA\ of $5.3 \times
10^{-17}$ erg s$^{-1}$ cm$^{-2}$ \AA$^{-1}$.  The relative fluxes
should be reliable at the 5\% level.  Figure 3 includes an earlier
spectrum on day 51 (green) where no evidence of dust is seen, and
broad-band {\it BVRIJHK} photometry from 4--5~d later, showing
early signs of the IR excess.  The $BVRI$ photometry is from our
database (see Paper I) and the $JHK$ measurements are from Arkharov et
al.\ (2006).  We include error bars for all of the broad-band photometry,
but the quoted uncertainty of 0.03 mag in the $JHK$ bands from
Arkharov et al.\ is smaller than the plotting symbols.

Photometry at longer IR wavelengths may constrain the dust emission
and chemistry better than our optical spectra, since we expect
$\lambda$F$_{\lambda}$ values of the hot dust to peak in the $H$ and
$K$ bands or at even longer wavelengths.  The graphite and silicate
models in Figure 3 predict fluxes extrapolated in the $H$ and $K$
bands that differ by 0.2 and 0.4 mag, respectively, with the weaker
fluxes corresponding to graphite.  These differences are large enough
that near-IR broad-band photometry should be able to constrain the
dust chemistry by comparison with Figure 3.  However, Figure 3 should
not necessarily be regarded as a {\it prediction} of the total IR
emission, but rather, only an extrapolation of emission from the
hottest newly formed dust.  It should therefore be regarded as a lower
limit to the near/mid-IR emission under each assumption for the grain
composition.  If our preferred scenario (see \S 6) is correct, we
actually expect cooler dust to ``pile up'' and give rise to greater IR
excess emission at longer mid-IR wavelengths, above the flux indicated
by the curves in Figure 3.

\begin{figure}
\epsscale{0.99}
\plotone{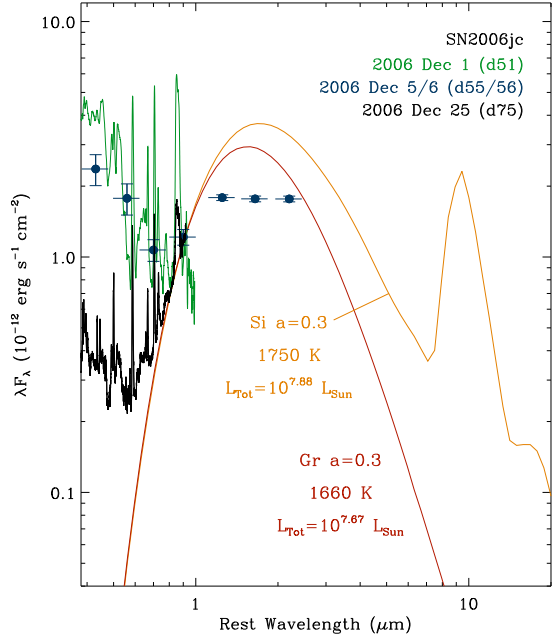}
\caption{The spectrum and dust models for 2006 Dec. 25 from Figure 2
  extrapolated into the IR.  The observed spectra were flux calibrated
  as noted in the text, and the models show predicted values of
  $\lambda$F$_{\lambda}$ for the silicate (orange) and graphite (red)
  grains from Figure 2.  The integrated luminosity of each dust
  component is noted as well.  The black spectrum is from day 75 when
  the IR emission appears to have peaked.  The day 51 spectrum from
  Figure 2 is shown in green, and the broad-band {\it BVRIJHK} spectral
  energy distribution from a few days later is shown in blue (see
  text).}
\end{figure}

The bolometric IR luminosity of the new dust that appears to have
reached its peak emission on Christmas Day 2006 (Fig.\ 3) is roughly
$5 \times 10^7$ L$_{\odot}$ for graphite or $8 \times 10^7$
L$_{\odot}$ for silicates (the difference is due to the different
wavelength-dependent emissivities of the two materials). This IR dust
luminosity is several times greater than the integrated optical
luminosity of the SN itself at the same epoch, estimated as follows.
If we extrapolate the late-time decline rate for the light curve in
Paper~I, as noted above, and we assume a distance modulus of 31.8 mag
for the host galaxy UGC~4904, we find that SN~2006jc should have rough
absolute {\it BVRI} magnitudes of $-$12.26, $-$12.48, $-$12.98, and
$-$13.83, respectively, on day 75.  This relative photometry should
also be accurate at the 5\% level.  If we adopt plausible bolometric
corrections of $-$0.74, $-$0.09, $-$0.32, and $-$0.66 mag for these
filters (see Paper~I), we find an average bolometric magnitude of
$-$12.85, or an intrinsic luminosity of roughly 10$^7$~L$_{\odot}$.
The fact that this is several times less than the integrated dust
luminosity agrees with the general impression of the
$\lambda$F$_{\lambda}$ plot in Figure 3, where the blue/near-UV
luminosity on day 75 is far below the peak of the IR emission.  This
luminosity difference supports the idea (see \S 6) that the dust does
not reside within the SN ejecta where it would be heated continually
by the central engine of $^{56}$Ni decay, because there is not enough
radiative luminosity to heat the dust.  Instead, the IR emission draws
its energy supply from the dense, post-shock radiative cooling zone.

The bolometric IR luminosity of the dust component can also be used to
estimate the mass of hot emitting dust; below we take graphite grains
as an example.  As long as the grains are small ($a\la$0.3~$\micron$),
the mass of emitting dust can be expressed independent of the assumed
grain radius and emissivity as

\begin{displaymath}
M_d = [(100 \rho) / (3 \sigma T_d^6)] \ \times \ L_d,
\end{displaymath}

\noindent 
where $\rho = 2.25$ g cm$^{-3}$ is the assumed density of graphite
grains, $\sigma$ is the Stefan-Boltzmann constant, $T_d$ is the
observed dust temperature of $\sim$1600~K, and $L_d$ is the total dust
luminosity (see Smith \& Gehrz 2005; Gehrz 1999; Gilman 1974).  From
the observed dust luminosity and temperature, we find a dust mass of
$\sim$6$\times$10$^{-6}$~M$_{\odot}$ present on day 75.  This is only
the mass of the very hottest emitting dust present at the time of the
observations; it is not the total mass of dust formed by SN~2006jc,
because this dust will cool rapidly and will be continuously replaced
by additional hot dust as long as the dust-formation epoch continues.
This can only be regarded as a minimum mass, since we are insensitive
to cooler dust emitting at longer IR wavelengths.

The total (i.e., cumulative) mass of dust produced as the shock sweeps
all the way through the dense shell is this instantaneous value
multiplied by $\Delta t$/$t_{\rm cool}$, where $\Delta t$ is the
observed duration of dust formation (roughly 1--2 months), and $t_{\rm
cool}$ is the cooling timescale of the dust in the dense post-shock
gas.  If the gas-grain cooling time is short (less than, say, 1 hr),
as we might expect for the observed temperatures and very high gas
densities in the post-shock shell (see Burke \& Hollenbach 1983), then
the total dust mass produced by SN~2006jc could be as high as 1\% of a
solar mass or more (and hence, the mass ejected in the LBV-like event
may have been as high as 1 $M_{\odot}$ for a normal gas:dust ratio).
This is only an order of magnitude estimate, of course, because of the
large uncertainty in the value of $t_{\rm cool}$.  Over time, we would
predict the following approximate behavior: As dust grains continued
to form over the 1--2 months when we detect the hot near-IR excess, a
range of dust temperatures will be seen.  The hottest newly formed
dust will continue to be seen at $\sim$1600~K, while dust that has
cooled will ``pile up'' at progressively lower temperatures, causing
the flux to rise at longer IR wavelengths.  After active dust
formation stops (sometime before day 128) this range of dust
temperatures will continue to cool as the dust expands and the SN
fades, and its emission will shift into the mid-IR.

Other changes also occurred in the spectrum of SN~2006jc at the same 
time as the dust formation, supporting the view that a substantial
amount of new dust formed.  Between days 50 and 75, the blue continuum
was seen to change its shape for the first time, having faded strongly
compared to the $R$-band flux (Fig.\ 1).  Moreover, starting at
around day 50, the visual light curve of SN~2006jc began to decline
faster than the $^{56}$Co decay rate (Paper~I).  Similarly, both the
Ca~{\sc ii} near-IR triplet and O~{\sc i} $\lambda$7774 faded away as
the red continuum excess grew stronger.  We speculate that the
fading of the blue continuum and the fading of these lines formed 
in the SN ejecta are both a consequence of obscuration by the
newly synthesized dust.  We discuss the possible effects of extinction
by dust in the next section.

The red excess emission seems to have disappeared by mid-February 2007
(day 128), although the SN was so faint by this time that our spectra
may be contaminated by light from an underlying star cluster and
H~{\sc ii} region.  This rapid disappearance of the dust emission is
unlike the case in SNe~IIn with dust, where the IR excess typically
lasts for years (Gerardy et al.\ 2002).  Did the red dust emission in
SN~2006jc vanish because the grains were destroyed or because they
cooled?  We offer a possible answer in the following section.

\begin{figure*}
\epsscale{0.9}
\plotone{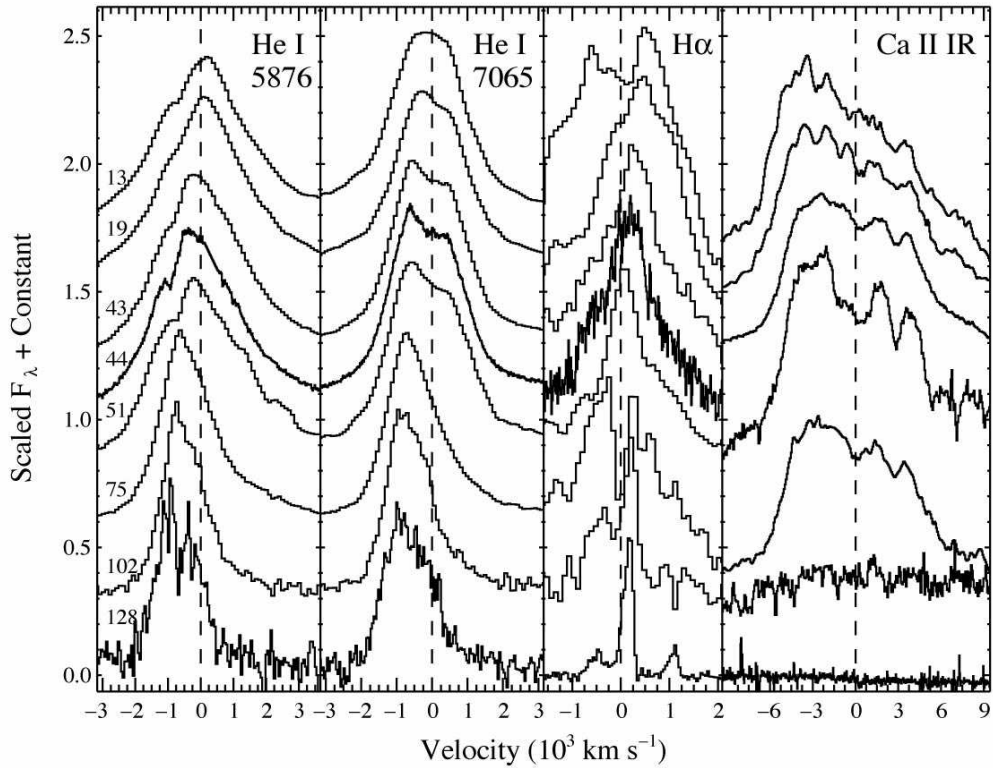}
\caption{Temporal evolution of the intermediate-width He~{\sc i}
  $\lambda$5876 and $\lambda$7065 emission-line profiles in SN~2006jc
  from early times (day 13) through our last spectrum in the nebular
  phase at day 128.  The He~{\sc i} lines show clear changes in their
  line profiles, becoming progressively more asymmetric and
  blueshifted, presumably due to increased obscuration from freshly
  synthesized dust.  H$\alpha$ and the broad Ca~{\sc ii} near-IR
  triplet, on the other hand, do not exhibit this effect.  H$\alpha$
  shows irregular profile changes, but they are not systematically
  shifted to the blue, while Ca~{\sc ii} simply fades entirely during
  the putative time of dust formation.  This gives an important clue
  to the location of the new dust (see text \S 5). (Note that in the
  last two spectra, the very narrow component of H$\alpha$ is probably
  from a superposed H~II region.  Also, the Ca~{\sc ii} profile is
  missing for day 44 because the spectrum on that date did not reach
  sufficiently long wavelengths.)}
\end{figure*}

\begin{figure}
\epsscale{0.95}
\plotone{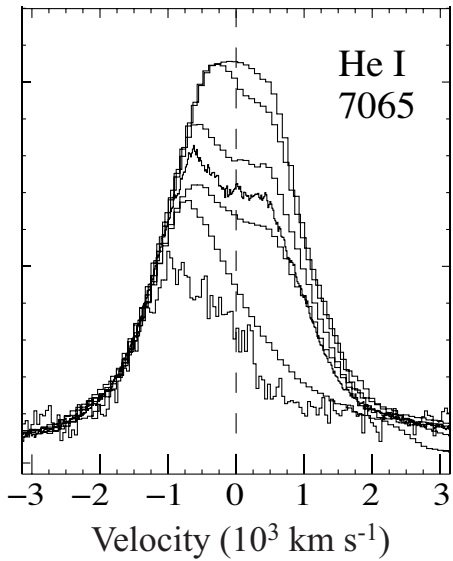}
\caption{Same as in Figure 4, but showing only He~{\sc i}
  $\lambda$7065 with all epochs scaled and superposed on one another
  to emphasize the change in line profile.}
\end{figure}


\section{EVIDENCE FOR DUST: CHANGING LINE PROFILES}

Another type of observation at optical wavelengths that has been
interpreted as evidence for dust formation in SN ejecta is the
development of asymmetric line profiles with a net blueshift as the
ejecta expand and cool.  Specifically, as dust grains condense in the
SN ejecta, one would expect this dust to obscure progressively more
emission from the far side of the object.  This should cause optically
thin emission-line profiles to become more asymmetric in shape and to
shift the bulk of their emission toward bluer velocities as emission
from the redshifted gas is blocked.  This effect has been documented
in several SNe~II, most notably in SN~1987A (Danziger et al.\ 1989;
Lucy et al.\ 1989; Colgan et al.\ 1994; Wang et al.\ 1996), indicating
the formation of new dust in the ejecta at $\sim$500~d after
explosion.

Whether this effect is common in SNe~Ib/c is unclear, as there are few
documented cases.  Elmhamdi et al.\ (2004) attributed the blueshifted
lines in the SN~Ib 1990I to condensation of dust in the ejecta around
day 250 after explosion.  In most cases, however, the asymmetric line
profiles in SNe~Ib/c have different interpretations.  Sollerman et
al.\ (1998), for example, argue that the blueshifted lines in the
ejecta of the SN~Ib 1996N were due to intrinsic asymmetry in the
ejecta and not the effect of dust obscuration described above.  The
blueshifted profiles in SN~1993J, which transitioned from Type II to
Type Ib, have been interpreted as a result of either clumping or optically
thick ejecta (Wang \& Hu 1994; Filippenko et al.\ 1994), but not dust
formation.

\begin{figure}
\epsscale{1.1}
\plotone{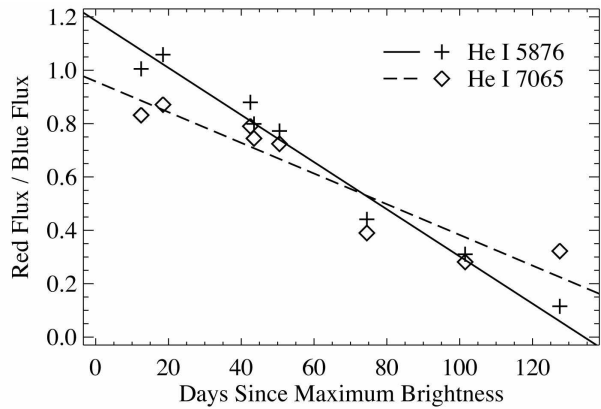}
\caption{The data in this plot document the fading of the red half of
  the intermediate-width He~{\sc i} lines with time, as compared to
  the blue side of the lines.  The plotted quantity is the ratio of
  the line flux at positive velocities to the flux at negative
  velocities, relative to the assumed systemic velocity shown with a
  dashed line in Fig.\ 4 and 5.  The straight lines show
  least-squares fits to the declining red/blue ratio of each line,
  with slopes of $-0.88 \pm 0.08$ per 100~d for $\lambda$5876 and
  $-0.58 \pm 0.09$ per 100~d for $\lambda$7065.  The last datum for
  $\lambda$5876 was not included in the fit because of contamination
  from Galactic Na~{\sc i}~D absorption, but this had little effect on
  the slope.  The fact that these slopes are significantly different
  is probably the effect of reddening from small grains, causing more
  severe extinction in $\lambda$5876 than in $\lambda$7065.}
\end{figure}

By contrast, we find very clear evidence in the spectra of SN~2006jc
that grain formation is affecting the line profiles as described
above, although the way it manifests itself is unusual compared to all
other SNe with this type of evidence for dust.  Specifically, the net
blueshift and asymmetry are seen primarily in the strong
intermediate-width He~{\sc i} emission lines that arise in the CSM or
post-shock shell (see \S 6.1), while the effect is not seen in broad
lines from the SN ejecta.

Figure 4 shows the line profiles of the two strongest intermediate-width 
He~{\sc i} lines in the optical spectrum of SN~2006jc as a function of 
time, while Figure 5 shows He~{\sc i} $\lambda$7065 alone, with all 
epochs aligned and plotted over one another for direct comparison of the 
line-profile evolution.  These lines, He~{\sc i} $\lambda$5876 and
$\lambda$7065, are relatively free from P-Cygni absorption, which
complicates the interpretation of He~{\sc i} lines superposed on the
strong blue continuum (Paper I).  Figures 4 and 5 display a clear
systematic evolution of the He~{\sc i} line profile with time.  Since
the lines appear nearly symmetric at early times, this effect is not
attributable simply to intrinsic asymmetries in the ejecta.  It also
cannot be due to optically thick ejecta as is thought to be the case
for SN~1993J (Wang \& Hu 1994; Filippenko et al.\ 1994), since that
interpretation would predict the lines to start very asymmetric and to
become progressively {\it less} asymmetric with time --- exactly the
opposite of what we observe in SN~2006jc. Instead, the increasing
asymmetry with time implicates dust formation in SN~2006jc.  The
He~{\sc i} profiles begin to show subtle effects of asymmetry around
days 40--50, but they become more severely asymmetric by day 75.
Notably, this epoch between day 51 and day 75 that shows the strongest
decrease in the red side of the lines is also the same time during
which the red/IR continuum appeared and when the blue continuum faded.

This severe asymmetry seen by day 75 persists through days 102 and
128.  By day 128 the red/IR continuum emission component had
disappeared (Fig.\ 2), while the obscuration of the red side of the
He~{\sc i} lines remained.  The newly formed dust was not destroyed;
its red emission simply faded because the dust cooled so that its
bolometric flux shifted to longer wavelengths, while its influence
through extinction persisted.

Figure 6 quantifies the time-dependent asymmetry of the He~{\sc i}
lines, where we show the integrated flux ratio of the red side of the
line to the blue side.  He~{\sc i} $\lambda$5876 and $\lambda$7065
exhibit slightly different behavior in the way their red flux
decreases with time.  Least-squares fits to the data indicate that the
asymmetric obscuration effect is more pronounced in He~{\sc i}
$\lambda$5876 than in $\lambda$7065, with statistically significant
differences in the slopes of their respective declining red/blue
ratios (Fig.\ 6).  Since $\lambda$5876 has a shorter wavelength than
the other, we might expect it to be more severely affected by dust
obscuration if the grains are small compared to the observed
wavelength.  This supports our assumptions about the grain size made
in \S 3 and Figure 3.  Of course, the trends in Figure 6 may not
follow straight lines.  In fact, the sharpest decrease appears between
days 50 and 75, as noted earlier, which is when the red/IR emission
strengthened.  The red side of the line decreases to roughly 40\% of
the emission on the blue side, indicating a visual-wavelength
extinction of roughly 1 mag toward the far side of the SN.

\section{RULING OUT LIGHT-ECHO MODELS}

The leading explanation for IR excess emission in SNe~IIn is a
light-echo model, where radiation from the SN propagates outward into
the pre-SN circumstellar ejecta.  In doing so it illuminates and heats
pre-existing dust formed in the progenitor's wind (Gerardy et al.\
2002; Fassia et al.\ 2000; Emmering \& Chevalier 1988; Dwek 1983; Bode
\& Evans 1980).

SN~2006jc is unique in that it was observed to have an LBV-like
eruption 2~yr before the SN, creating a dense environment.  Normal LBV
eruptions are known to be copious dust producers (e.g., Smith et
al. 2003), so we cannot exclude the possibility of considerable dust
in the CSM around SN~2006jc.  Thus, we consider whether an IR echo can
explain our observations of SN~2006jc, but we find that it cannot for
two main reasons.

First, the dust around SN~2006jc was much hotter ($\sim$1600~K) than
that in SNe~II (a few hundred K).  It stayed at a roughly constant
temperature until it disappeared quickly, whereas IR light echoes tend
to decay slowly on timescales of years (Gerardy et al.\ 2002).  These
hints point toward newly formed dust rather than dust at large radii
that is radiatively heated by a central engine.

Second, the evolution of the He~{\sc i} line profiles indicates an
amount of obscuring dust that increases systematically with time.  In
an IR-echo model, the amount of dust and the extinction it causes stay
constant with time or may even decrease.  Furthermore, such dust would
be at large radii and would not selectively absorb red sides of the
line profiles.  Thus, we can rule out an IR echo as the principal
agent for the dust seen in SN~2006jc.

\section{WHERE IS THE NEW DUST LOCATED?}

Sometime between days 51 and 75, several events occurred
simultaneously in the spectroscopic evolution of SN~2006jc which are
relevant to the interpretation of dust formation in this object.

(a) SN~2006jc developed a strong red/IR continuum excess emission
    component, which can be accounted for by hot graphite grains at
    around 1600~K, or silicates at slightly higher temperatures.  In
    both cases, the grains are likely to be smaller than the observed
    wavelength.

(b)  The intermediate-width He~{\sc i} line profiles became severely 
    asymmetric and
    blueshifted, suggesting that the redshifted emitting material
    has been largely obscured.  This effect was more pronounced at
    shorter wavelengths, again pointing to relatively small grain radii.

(c) The strong blue continuum faded relative to the red continuum for
    the first time, and the broad emission lines tracing the SN ejecta
    (like Ca~{\sc ii} and O~{\sc i}) faded.  Unlike the He~{\sc i}
    lines, the broad SN ejecta emission lines did not show obvious
    blueshifts or more asymmetric profiles --- they simply faded (Fig.\
    4).  At the same time, the $B$-band light curve started to fade
    faster than the $^{56}$Co decay rate (Paper~I).

\smallskip

The details of how these changes happened help to unravel the
geometry of the SN ejecta.  In particular, they provide clues to
the relative locations of the obscuring and emitting dust, the 
intermediate-width He~{\sc i} lines, the blue continuum, and the 
broad emission lines.

\begin{figure}
\epsscale{0.95}
\plotone{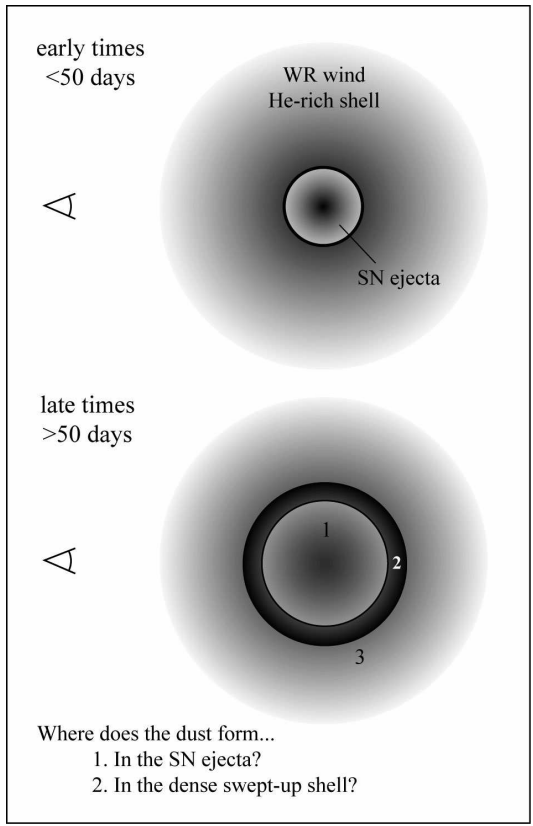}
\caption{Sketch of a simplified SN~2006jc geometry for the purpose of
  discussing where the dust forms.  The top sketch is at early times,
  when the SN blast wave has started plowing into the He-rich CSM, but
  has not yet swept up much mass.  The bottom sketch is at later
  times, closer to the epoch of dust formation, when a thicker, dense
  shell has developed in the cooling zone behind the forward shock.
  The two locations where dust is likely to form are zone 1 in the SN
  ejecta themselves as they cool enough to allow condensation (this is
  the same zone that gives rise to the broad Ca~{\sc ii} and O~{\sc i}
  emission), or zone 2 in the dense shell of swept-up gas between the
  reverse shock and the forward shock. There are geometrical reasons
  to favor zone 2 as the most likely region of dust formation.
  Intermediate-width He~{\sc i} lines arise in zone 2 and/or 3, close
  to the blast wave where the densities are high.}
\end{figure}

Figure 7 shows a cartoon of the possible geometry of SN~2006jc,
limited to the structures relevant for interpreting effects of dust.
We assume spherical symmetry for simplicity, but the discussion below
does not preclude possible asymmetry in the SN ejecta or the bipolar
geometry that is likely in circumstellar matter arising from LBV-style
eruptions (e.g., Smith 2006; Owocki 2003).  We can estimate the outer
radius of the dense He shell ejected in the 2004 event as $v_W \times
\Delta t \approx 2000 \ {\rm km \ s^{-1}} \times 2 {\rm yr} \approx
1000$~AU.  The radius of the blast wave is $v_{\rm shock} \times
\Delta t$, so for a shock speed of $\sim10^4$ km s$^{-1}$, the
timescale for the shock to sweep through most of the He shell is
$\sim$100--200~d.  This is comparable to the timescale of our
observations, so we might expect significant changes as swept-up
material accumulates.

\subsection{The Emitting He~{\sc i} Region}

Before determining the zone in Figure 7 where the dust formation
occurs, we must first understand the origin of the bright
intermediate-width He~{\sc i} lines, potentially coming from the
pre-shock WR wind or the post-shock shell.  Both the wind or
post-shock origin are plausible, in principle, since the progenitor
star was thought to be a WR star, and such stars have fast winds
capable of producing the observed He~{\sc i} linewidths.  Figure~8$a$
shows the time evolution of line fluxes for He~{\sc i} lines in
SN~2006jc measured from our spectra.

\begin{figure}
\epsscale{0.98}
\plotone{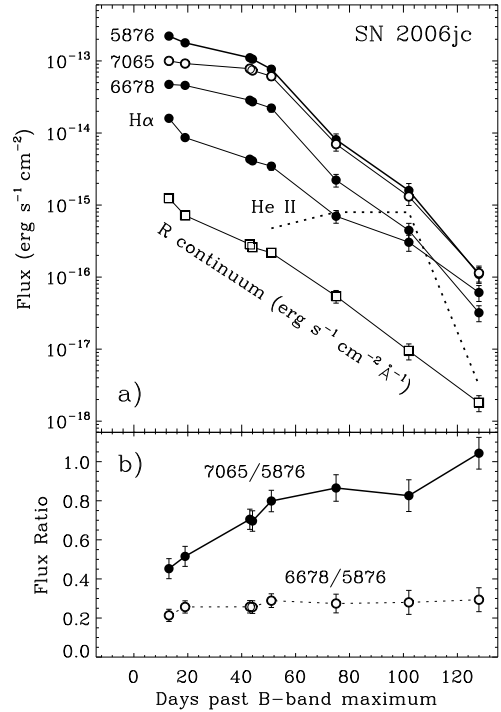}
\caption{Time evolution of emission-line fluxes in SN~2006jc.  Panel
  (a) shows the measured flux for bright He~{\sc i} lines and
  H$\alpha$, compared to the decline rate of the red ($\sim$6500~\AA)
  continuum (in $F_{\lambda}$ units) on the same dates as the
  emission-line measurements, interpolated and extrapolated from
  $R$-band photometry from Paper~I as described in the text.  The
  dashed curve shows the measured line flux of He~{\sc ii}
  $\lambda$4686 on days 75 and 102, where the first and last values on
  days 51 and 128, respectively, are upper limits (the uncertainty in
  the He~{\sc ii} fluxes on days 75 and 102 are about 10\%).  Panel
  (b) shows the observed ratios of He~{\sc i} $\lambda$7065 (solid
  line, filled circles) and $\lambda$6678 (dashed, unfilled) to
  $\lambda$5876 on the same dates.}
\end{figure}

In general, the emission-line fluxes decline with time at a rate
commensurate with the drop in the red continuum level.  The red
continuum in Figure 8$a$ represents the assumed continuum at the time
of each line measurement, interpolated from the $R$-band light curve in
Paper~I.  As noted earlier, we extrapolated the decline rate to infer
the red continuum flux at the latest epochs.  Since we calculated line
fluxes using these continuum values, uncertainties in line fluxes are
large at late times (we adopt generous 10--20\% error bars in Fig.\ 8).
However, since fluxes for various lines measured in the same
spectrum at each epoch are scaled using the same continuum,
the uncertainty in {\it relative} line fluxes is much smaller.

Of particular interest is He~{\sc i} $\lambda$7065, which seems to
counter the trend exhibited by the other two He~{\sc i} lines in
Figure~8$a$.  Namely, it has a slower decline rate, becoming more
prominent in the spectrum with time.

More telling is Figure 8$b$, which shows the time evolution of the
He~{\sc i} $\lambda$7065/$\lambda$5876 flux ratio.  These are the two
brightest lines in the spectrum, so measurement uncertainties are
small.  We see that He~{\sc i} $\lambda$7065 is only half as strong as
$\lambda$5876 at early times, and continually grows until the lines
have equal flux at late times.  This effect is not due simply to
wavelength-dependent extinction by dust, since it is not seen in the
He~{\sc i} $\lambda$6678/$\lambda$5876 flux ratio, also shown in
Figure 8$b$.

If the He~{\sc i} lines arise primarily in the CSM, then their flux
should drop with time as observed because they result from
photoexcitation by the SN light.  However, in this scenario it is
difficult to understand the growing He~{\sc i}
$\lambda$7065/$\lambda$5876 ratio, which is a hallmark of increasing
density, because the density of the CSM should drop as $r^{-2}$.

On the other hand, if the He~{\sc i} lines arise primarily in the
post-shock gas it could explain the high densities inferred from the
$\lambda$7065/$\lambda$5876 ratio, but we might expect their flux to
be constant with time or even to increase as more material enters
the post-shock shell.

To explain the evolution of relative He~{\sc i} line strengths, then,
we propose a hybrid scenario, where the observed He~{\sc i} lines are
emitted from {\it both} the unshocked CSM and the post-shock shell ---
but where their relative contributions to the spectrum change with
time.  At early times (upper sketch in Fig.\ 7), the emission from the
unshocked CSM has the advantage of a large emitting volume, so it
dominates the spectrum.  At late times, the post-shock emission takes
over for two basic reasons. First, the shock has swept through much of
the CSM volume and placed more of the He gas in the post-shock region,
detracting from the CSM emission that is dropping anyway because of
the decline in UV radiation.  Second, the much higher density of
post-shock gas ($n_e \approx 10^{10}$ cm$^{-3}$ for
$F$($\lambda$7065)/$F$($\lambda$5876) $\ga$ 1; see \S 6.3) has the
advantage that emission is proportional to $n_e^2$, making it
increasingly dominant with time as more material passes through the
shock.  In this model, we might expect the He~{\sc i} line widths to
increase with time, but if this effect is present, it is hidden by
growing extinction from dust, making the lines asymmetric (Fig.\ 5).

Lastly, we note that Figure 8$a$ also shows the evolution of the
H$\alpha$ line flux with time, indicating that it too grows stronger
compared to He~{\sc i} $\lambda$5876.  However, this increase in the
relative strength of H$\alpha$ is most likely due simply to increased
contamination from a background H~{\sc ii} region.  Since the H~{\sc
ii} region will be constant with time, it will contribute relatively
more to the spectrum as the SN light fades.  At late times, the
H$\alpha$ line is dominated by a very narrow (unresolved) line with
neighboring [N~{\sc ii}] lines typical of H~{\sc ii} regions.  The
intermediate-width component of H$\alpha$ is barely visible at late
times, contributing less than 5\% of the total flux.  By contrast,
such narrow (unresolved) components for He~{\sc i} are not seen.

\subsection{Zones 1, 2, and 3}

At early times shown in the top panel of Figure 7, the SN ejecta
occupy a small volume compared to the He-rich WR wind, and the blast
wave has swept up relatively little mass.  The top panel corresponds
to the early epochs of spectra discussed in Paper~I.  The late-time
spectra discussed here correspond to the lower panel in Figure 7, when
the central SN ejecta (zone 1) have faded and when a significant
fraction of the mass in the CSM (zone 3) has been swept up to form
part of a dense shell between the forward and reverse shocks (zone 2).
The broad emission lines like the Ca~{\sc ii} near-IR triplet and
O~{\sc i} $\lambda$7774 originate in zone 1, while the
intermediate-width He~{\sc i} lines originate in zone 3, the outer
part of zone 2, or both.  Again, if the progenitor was a WR star, its
likely wind speeds of 1000--2000 km s$^{-1}$ are sufficient to explain
the intermediate-width lines.  The strong blue/UV continuum also
originates in zone 1, because blue He~I lines show P~Cygni absorption
of the blue continuum, whereas red He~{\sc i} lines do not (Paper I).
Since there are singlet and triplet He~{\sc i} lines in both
wavelength ranges, the likely reason for the presence of P~Cygni
absorption in the blue and not in the red is probably related to the
geometry or the strength of the blue continuum.

Given the hypothetical geometry in Figure 7, consider the
observational consequences for the two likely locations where
the dust in question may have formed, which are in the SN ejecta (zone
1) or in the dense swept-up shell (zone 2):

(1)  If dust grains form in the SN ejecta in zone 1, they can
    only block part of the total He~{\sc i} emission from
    the observer, and would therefore only influence the directly receding
    material in the red wings of the He~{\sc i} lines.  Dust grains in
    this central region would not be able to absorb any of the He~{\sc i} 
    emission near the systemic velocity, and therefore cannot
    account for the observed He~{\sc i} line profiles from day 75
    onward, when most of the red emission and some of the
    zero-velocity emission is absorbed.  Additionally, dust forming
    within the SN ejecta would be expected to preferentially absorb
    emission from redshifted SN ejecta, leading to asymmetric Ca~{\sc
    ii} and O~{\sc i} line profiles.  These are not observed (Fig.\
    4), so we conclude that it is unlikely that the dust has formed in
    the SN ejecta in zone 1.

(2) If instead the dust grains form in the dense swept-up shell in
    zone 2, then they can efficiently prevent much of the He~{\sc i}
    emission from reaching the observer, because He~{\sc i} lines
    are formed in a region with a similar extent.  This could explain
    the line profiles in Figures 4 and 5.  Also, dust formed in zone 2
    would absorb emission from the SN ejecta equally at all
    velocities, consistent with the lack of blueshifted profiles in
    the broad Ca~{\sc ii} and O~{\sc i} lines.  Since the blue
    continuum probably arises in zone 1, dust formed in zone 2 could
    also explain its fading by day 75 (Fig.\ 1).  For these
    reasons, we consider it more likely that grains are
    condensing in the swept-up shell in zone 2.

\subsection{Steady-State Dust Formation in the Post-Shock Gas}

At first glance, it seems somewhat odd that the apparent grain
temperature around 1600~K stayed nearly constant (Fig.\ 2) for about a
month after Christmas 2006, and then after another month (by
mid-February 2007) the dust had cooled rapidly enough to be invisible
at optical wavelengths.  This would be difficult to understand in a
scenario where the grains condense directly in the SN ejecta (zone 1
in Fig.\ 7) and are warmed by a declining luminosity from radioactive
decay; in fact, this interpretation seems impossible because the SN
ejecta have insufficient luminosity to heat the dust (see \S 3.2).
However, if the grains are forming in the dense swept-up shell (zone 2
in Fig.\ 7) as we suspect, there may be a reasonable explanation as
follows.

When a blast wave encounters and sweeps through a dense circumstellar
shell, we can hypothesize that grains may continuously form at $T
\approx 1600$~K as dense material rapidly cools behind the shock
front.  This may continue as long as sufficiently dense matter enters
the shock.  How dense is the post-shock gas?  As noted earlier, we
measure an unusually high flux ratio of He~{\sc i}
$\lambda$7065/$\lambda$5876, approaching unity (Fig.\ 8$b$).  This was
seen in SN~1999cq as well (Matheson et al.\ 2000).  Following Almog \&
Netzer (1989), this would seem to indicate electron densities of
almost $10^{10}$ cm$^{-3}$ (although their calculations were for He
embedded in H-rich gas), or mass densities of $\sim 3 \times 10^{-14}$
g cm$^{-3}$ for ionized He gas.  Coincidentally, this value is
compatible with the expected critical density for graphite dust
precipitation (e.g., Clayton 1979).  Such high densities are far too
high to exist in the CSM, because they would require mass-loss rates
greater than 10~M$_{\odot}$ yr$^{-1}$ at $R = 500$ AU for a wind speed
of $\sim$2000 km s$^{-1}$, and would obscure our sightline to the SN.

Instead, these densities are only likely to be found in the post-shock
cooling zone where material piles up.  Coincidentally, efficient
He~{\sc ii} $\lambda$4686 emission, which is correlated with the dust
emission as discussed next in \S 7, also requires densities of order
$10^{10}$ cm$^{-3}$ (e.g., Martin et al.\ 2006).  From our
observations alone we cannot differentiate between dust forming in CSM
gas swept up by the forward shock, or instead, C-rich SN ejecta that
have passed the reverse shock.  However, the overlap with the
conditions leading to dust formation in the C-rich colliding winds of
WC+O binaries described in \S 7 would add weight to the latter
option in the case of SN~2006jc.  Detailed models of dust formation in
dense shocks would be interesting to pursue in this regard.

The hot dust that forms will cool rapidly and its optical emission
will fade, but it will continuously be replenished as long as the
shock is sweeping up sufficiently dense matter.  Thus, the freshly
synthesized dust at the hottest temperatures will always dominate the
flux at the shortest wavelengths that we observe here.  This
steady-state phase may continue until the shock passes beyond the
densest part of the LBV-like shell, when the density drops below the
critical threshold for grain nucleation.  Optical spectra of SN~2006jc
in Figure 2 suggest that this phase lasted about $\Delta t \ga 1$
month, implying a thickness for the hypothetical shell of roughly
50--200 AU for blast wave expansion speeds of 2,000--10,000
km~s$^{-1}$.  Perhaps it is only a coincidence that the radial
thickness of the massive dust shell around $\eta$~Carinae is a few
hundred AU as well (Smith 2006).  In any case, the 1-month-long
constant-temperature phase seen in the red continuum feature (Fig.\ 2)
suggests that its {\it emitting} dust is consistent with a scenario
involving dust formation in a dense swept-up shell, like the {\it
obscuring} dust discussed above.

\begin{figure}
\epsscale{0.95}
\plotone{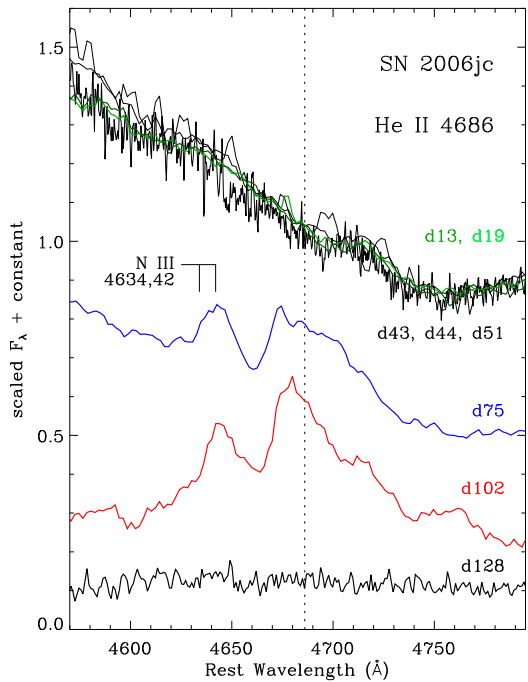}
\caption{Similar to Fig.\ 1, but emphasizing the wavelength range
  around He~{\sc ii} $\lambda$4686.  The early-time spectra of days 13
  through 51 show no evidence for an emission feature that can be
  associated with He~{\sc ii}.  A prominent, intermediate-width 
  (FWHM $\approx$
  2000 km~s$^{-1}$) emission feature that could be He~{\sc ii}
  $\lambda$4686 (dashed vertical line) with an irregular line profile
  is seen at day 75, strengthens by day 102, and then disappears
  completely by day 128.  Thus, the He~{\sc ii} emission feature
  follows the same temporal behavior as the red continuum from dust in
  Figure 1.  The relatively narrow emission feature on the blue side 
  of the line is probably N~{\sc iii} $\lambda$4640.}
\end{figure}

\section{H\lowercase{e}~{\sc ii} $\lambda$4686, DUST FORMATION, AND ETA CARINAE}

The CSM that is struck and swept up by the forward shock will be a
hot, X-ray--emitting plasma that must cool efficiently in order to
form dust grains.  Understanding this process will require detailed
calculations of the shock interaction that are well beyond the scope
of our paper.  Nevertheless, we argue that this scenario is plausible
because it has a related observational precedent: dust formation is
known to occur in the strong colliding-wind shocks of WC+O binaries
(e.g., Allen et al.\ 1972; Gehrz \& Hackwell 1974; see also Crowther
2003 for a review).  How dust is able to form in these inhospitable
conditions is not fully understood either, but clearly nature is able
to overcome this obstacle in the compressed post-shock gas because the
evidence for dust formation in these systems is unambiguous.

Prototypes of the dust-forming late-type WC binaries are the
persistent dust producers like WR104 and WR98a (the so-called
``pinwheel'' systems; Tuthill et al.\ 1999; Monnier et al.\ 1999) and
the episodic dust producers in eccentric binaries like WR140 (Monnier
et al.\ 2002).  In such systems, strong colliding-wind shocks compress
and heat He-rich and C-rich gas to temperatures capable of emitting
strong X-rays (e.g., Usov 1999).  As the gas proceeds down the shock
cone, it then cools rapidly enough to produce graphite dust before
dispersing.  These conditions closely mirror the shocked shell (zone 2
in Fig.\ 7) of SN~2006jc, which is also an X-ray source (Immler et
al.\ 2008).  Thus, in the case of SN~2006jc, the C-rich SN ejecta
passing through the reverse shock might provide the necessary seeds
for grain nucleation.  As noted earlier, another clue that the dust
may be C-rich is that it appears to condense at a temperature of
$\sim$1600~K, higher than the temperatures at which silicate grains
are normally seen to condense.

Another eccentric colliding-wind binary that looks almost identical to
WR140 in X-rays is $\eta$ Carinae (see, e.g., Corcoran 2005 for a
review and references), where X-ray emission is thought to result from
the collision of a dense and relatively slow LBV wind with a faster
$\sim$3,000 km s$^{-1}$ O-star wind (thus, the characteristic shock
velocities may be similar to those of SN~2006jc).  In $\eta$ Carinae, X-ray
emission is relatively weak for most of the $\sim$5 yr orbit, but the
strength of X-rays greatly intensifies as the system approaches
periastron when the post-shock density increases dramatically.  At the
same time, strong near-IR excess emission develops and continues to
rise toward periastron along with the X-ray emission (e.g., Whitelock
et al.\ 2004).  The coincidence of X-ray emission and near-IR excess
emission is reminiscent of SN~2006jc, although in $\eta$ Car it is not
known if the near-IR excess is due to newly formed hot dust grains in
shocks.  In $\eta$ Carinae, the rise of X-rays and near-IR emission
around periastron is also accompanied by a brief occurrence of He~{\sc
ii} $\lambda$4686 emission that is weak or absent from the spectrum at
all other times (Steiner \& Damineli 2004; Martin et al.\ 2006).  This
may hint at a link of common physical conditions that are required to
produce X-rays, a near-IR excess, and He~{\sc ii} emission in dense
shocks.

In that case, the spectra of SN~2006jc in Figure 8 become quite
intriguing.  An emission feature that can plausibly be identified as
He~{\sc ii} $\lambda$4686 is seen in the spectrum.  It is an
intermediate-width feature much like the He~{\sc i} lines, spanning
roughly $\pm$2500 km~s$^{-1}$.  It has an irregular double-peaked
shape because it is accompanied by N~{\sc iii}
$\lambda\lambda$4634,4642 emission in its blue wing that directly
varies with the strength of He~{\sc ii} $\lambda$4686.  Following
comments in \S~1, it is curious that these features are
also defining characteristics of the dense winds in Of/WN stars.

Figure 8$a$ shows the He~{\sc ii} $\lambda$4686 flux as a function of
time with a dotted line, where the first and last points are upper
limits because the line was only detected on days 75 and 102.  Days 75
and 102 showed nearly identical He~{\sc ii} $\lambda$4686 line fluxes
of $\sim$10$^{-15}$ erg s$^{-1}$ cm$^{-2}$ (excluding N~{\sc iii}),
translating to an intrinsic luminosity of $1.3 \times 10^4$
L$_{\odot}$ if we adopt a distance modulus $m-M = 31.8$ mag as before.
The 4686~\AA\ line probably represents about 1\% of the cooling by all
He~{\sc ii} recombination lines.  If so, then He~{\sc ii}
recombination contributes $\sim$10\% of the UV/visual luminosity of
the SN.  Thus, either photoexcitation or shock energy could power the
He~{\sc ii} lines.

The important point, though, is that the feature around 4686~\AA\ is
seen {\it only} in the day 75 and day 102 spectra --- precisely the
same epochs when the strong red/IR continuum emission from dust is
present (Fig.\ 1--3).  It is completely absent in the early-time
spectra, and it disappears by the time of our last spectrum on day
128, just like the hot-dust emission.  This coincidence can be taken
as further supporting evidence that the dust formation in SN~2006jc
occurs in the post-shock gas (zone 2), where the He~{\sc i} and
He~{\sc ii} emission arise, analogous to the dust formation in
colliding-wind binaries.

Since He~{\sc ii} $\lambda$4686 and the IR excess emission from dust
seem to track one another in SN~2006jc and are apparently associated
with a strong shock event, it seems reasonable to predict that the
X-rays from SN~2006jc may behave similarly with time.  In fact, Immler
et al.\ (2008) recently noted a distinct and very unusual increase in
the X-ray flux of SN~2006jc that lasted for a few months and coincided
precisely in time with our proposed epoch of dust formation and
enhanced He~{\sc ii} $\lambda$4686 emission in Figure 8$a$.  This
coincidence suggests a common link.

What does this imply for $\eta$ Carinae, which shows nearly identical
behavior of X-rays, He~{\sc ii} $\lambda$4686 emission, and near-IR
excess emission appearing for only a short time?  In SN~2006jc, the
case for newly formed dust grains in the post-shock gas seems solid.
We conjecture, then, that the near-IR excess in $\eta$ Carinae is also
the result of emission from hot ($\sim$1600~K) dust grains that have
formed in the dense post-shock gas in the colliding wind interaction
region where gas can evidently cool rapidly enough.  After this paper
was submitted for publication, we learned that Kashi \& Soker (2007)
independently came to a similar conclusion.

\section{SUMMARY}

Based on the red/near-IR continuum excess emission that can be
explained by hot dust at 1600--1700~K, as well as the systematically
more asymmetric and blueshifted He~{\sc i} emission lines, we propose
that SN~2006jc began to form significant amounts of hot dust between
50 and 75~d after peak luminosity.  This dust-formation epoch
continued for at least a month or more, and then apparently ceased.
Based on this behavior, a number of geometric considerations, and the
transient He~{\sc ii} $\lambda$4686 emission, we argue that the dust
formed in a dense shell of shocked gas and not in the SN ejecta. Grain
formation in this post-shock shell may also help account for why the
dust formed so soon after explosion --- sooner than in any other
dust-forming SN.  The shell could have been composed of dense CSM
ejected by an LBV-like event observed 2~yr prior to the SN (Paper~I;
Pastorello et al.\ 2007), which was then swept up by the forward
shock.  Alternatively, grains may have condensed from C-rich SN ejecta
that passed the reverse shock when the blast wave decelerated upon
impact with the LBV-like shell.  Either hypothesis would be consistent
with our observational constraints that place the dust somewhere
between the forward and reverse shocks (zone 2 in Fig.\ 7).  The
dust-formation epoch lasted for a brief phase when the swept-up
material behind the shock was sufficiently dense and cooling quickly
enough for refractory elements to condense.

Few other SNe~Ib/c have shown any evidence for dust, and SN~2006jc may
now provide the least ambiguous case where new dust has formed.  We
suspect that this --- like most of the other unusual properties of
SN~2006jc --- is a direct consequence of the dense circumstellar
environment around SN~2006jc ejected during the LBV-like eruption just
before explosion.  One obvious speculation is that SN~1999cq and
SN~2002ao, which had similar spectra to SN~2006jc (see Paper~I), may
also have undergone episodic mass ejection shortly before they
exploded.  SN~2006jc, SN~2002ao, and especially SN~1999cq had very
rapidly declining optical light curves compared to most other SNe~Ib/c
(Paper I).  Although the dust effects discussed in this paper were not
documented for either SN~2002ao or SN~1999cq, their rapid declines may
have been accompanied by dust formation.  It will be interesting to
see if such a connection between LBV-like precursor events and
SNe~Ib/c with intermediate-width lines can be established with a
future example.

Given the rarity of dust formation in SNe~Ib/c and the scarcity of
dense circumstellar environments like those of SNe~2006jc, 1999cq, and
2002ao, one might conclude that most SNe~Ib/c are {\it not} preceded
by an LBV-like eruption, and that SN~2006jc is therefore a special case.

The presence of weak hydrogen emission in the spectrum, and the
occurrence of an LBV-like outburst just before the SN, would seem to
imply that SN~2006jc was a young WN star that was still in the last
phases of transitioning from the LBV phase (Paper~I).  This seems at
odds, though, with the fact that SN~2006jc was basically a Type Ic
event with an external He-rich CSM, implying that it also recently
removed its He layer in an abrupt transition to the WC phase.  (No WC
stars are known to contain H, while some WN stars do; see Smith \&
Conti 2008.)  Even if the ejected layer had retained some H because of
previous mixing, that mixing must not have reached into deeper layers.
Alternatively, Pastorello et al.\ (2007) proposed that the mystery of
the CSM of SN~2006jc might be explained by invoking a less-evolved
companion star in the system that had a classical LBV eruption.
However, transferring the mysterious LBV event to a hypothetical
companion still does not account for why the CSM was H poor. In any
case, this overlap is reminiscent of SNe that blur the distinction
between Types~Ib and II, like SN 1987K (Filippenko 1988) and 
SN~1993J (e.g., Filippenko et al. 1993), and is mirrored by the 
ambiguity between LBVs and H-rich WN stars noted in \S~1.

Altogether, then, these clues suggest that rapid evolution through
episodic LBV-type mass-loss events may play a significant role in both
the onset and the progression of the WR phase. Just as giant LBV
eruptions may drive the evolution from H-rich to H-poor stars (Smith
\& Owocki 2006), perhaps similar events in WR stars can, in rare
cases, drive the evolution from WN to WC as suggested by SN~2006jc.
In this context, it is worth emphasizing that if the progenitor of
SN~2006jc had undergone core collapse only 5--10 yr ago, it would have
probably appeared as a relatively normal Type~Ib event.

\acknowledgments 
\footnotesize

We gratefully acknowledge discussions with R.\ Chornock, M.\
Ganeshalingam, W.\ Li, M.\ Modjaz, K.\ Nomoto, T.\ Nozawa, and C.\
Fransson.  Most of the data presented herein were obtained at the
W.\ M.\ Keck Observatory, which is operated as a scientific partnership
among the California Institute of Technology, the University of
California, and the National Aeronautics and Space Administration. The
Observatory was made possible by the generous financial support of the
W.\ M.\ Keck Foundation.  We thank the Keck and Lick Observatory staffs
for their assistance. This research was supported by NSF grant
AST--0607485.  Support for this work was also provided by NASA through
awards (Programs 20256 and 30292) issued by JPL/Caltech.


\end{document}